\documentclass[11pt]{article}

\usepackage{microtype}
\usepackage{booktabs}
\usepackage{url}
\usepackage{csquotes}
 \usepackage{multirow}

\usepackage{amsmath}
\usepackage{amsthm}
\usepackage{tabularx}
\usepackage{graphicx}

\usepackage[preprint]{automl}

\usepackage{natbib}
\title{Graph Instance Landscapes: When Structural Similarity Does (Not) Reflect Shortest-Path Performance}
 
\author[1,2]{\nameemail{Maryam Gholami Shiri}{maryam.gholami.shiri@ijs.si}}
\author[3]{\nameemail{Ivana Krminac}{ivana.krminac@student.pmf.unibl.org}}
\author[4,5]{\nameemail{Marko Djukanovi\'c}{marko.dukanovic@ung.si}}
\author[1]{\nameemail{Sa\v{s}o D\v{z}eroski}{saso.dzeroski@ijs.si}}
\author[1,6]{\nameemail{Eva Tuba}{eva.tuba@ijs.si}}
\author[1,2]{\nameemail{Tome Eftimov}{tome.eftimov@ijs.si}}
 
\affil[1]{Jo\v{z}ef Stefan Institute, Ljubljana, Slovenia}
\affil[2]{Jo\v{z}ef Stefan International Postgraduate School, Ljubljana, Slovenia}
\affil[3]{Faculty of Natural Science and Mathematics, University of Banja Luka, Banja Luka, Bosnia and Herzegovina}
\affil[4]{University of Nova Gorica, Nova Gorica, Slovenia}
\affil[5]{Institute of Information Sciences (IZUM), Maribor, Slovenia}
\affil[6]{Trinity University, San Antonio, TX, USA}

\begin{document}
\maketitle
 
\begin{abstract}
Benchmarking shortest-path algorithms is commonly based on aggregate performance over heterogeneous graph sets, which limits insight into how different search paradigms react to instance structure. We adopt an instance-landscape view of graph benchmarking by embedding graphs into a low-cost structural feature space and clustering them into regions of similar structure. Three benchmark suites are studied: weighted Erd\H{o}s--R\'enyi graphs, random geometric (wireless) graphs, and real-world road networks. We evaluate four representative shortest-path solvers spanning uninformed exact search (Dijkstra), bidirectional exact search (bidirectional Dijkstra), heuristic-guided exact search (A$^{*}$), and deque-based strategies (DEQ). Clustering robustness is analyzed under multiple feature-selection schemes, and runtime distributions are compared across landscape regions using non-parametric tests. While generator parameters induce stable structural regions, we find that feature-space similarity does not necessarily imply performance similarity: significant runtime shifts are frequently observed even within the same landscape region. A merged-suite analysis further shows that different benchmark families occupy largely disjoint regions. These results highlight both the potential and the limits of structural landscapes for the structure-aware benchmarking of shortest-path algorithms.
\end{abstract}
 
\section{Introduction}
\label{introduction}
Graph-based optimization problems play a central role in numerous real-world applications, including transportation systems, communication networks, logistics, and geographic information systems. Among these, shortest-path problems form a fundamental class that has been extensively studied, resulting in a wide range of exact algorithms such as Dijkstra's algorithm, the Bellman–Ford algorithm, A$^*$ search, and many of their variants and extensions~\citep{dijkstra2022note,bellman1958routing,cormen2022introduction,johnson1977efficient,bertsekas1993simple,guo2018improved}. These algorithms have proven highly effective in addressing a variety of practical problems~\citep{chakaravarthy2016scalable,federickson1987fast,sommer2014shortest,hadi2025comprehensive,pramudita2019shortest}.
 
Despite polynomial-time guarantees, it is well known that the practical performance of graph algorithms varies significantly across instances, depending on structural properties such as graph size, density, degree distribution, edge-weights distribution, and topology. As a consequence, extensive benchmarking studies have been conducted to compare algorithmic behavior across synthetic and real-world graph datasets~\citep{goldberg2005computing,sanders2006engineering}. These studies have shown that no single algorithm consistently dominates across all graph types and scales.
 
Most existing benchmarking efforts, however, focus on instance-level performance comparisons, typically reporting average runtimes, rankings, or statistical significance tests across predefined benchmark sets. While informative, such analyses provide limited insight into how the space of graph instances itself is structured and how systematic changes in graph generation parameters influence algorithmic behavior.
 
Recent work has highlighted the importance of instance features for understanding and predicting algorithm performance. By extracting structural characteristics from graphs and employing machine learning or explainable AI techniques, it is possible to identify the features that most strongly influence the efficiency of the algorithm~\citep{lundberg2017unified,smith2012measuring}. Nevertheless, these approaches usually analyze feature–performance relationships directly, without explicitly modeling the geometry of the instance space induced by different benchmark generators.
 
In this paper, we adopt a landscape-based perspective on graph benchmarking, inspired by instance space analysis in optimization. We represent graph instances as points in a feature space and apply clustering techniques to identify regions corresponding to distinct graph landscapes. This allows us to study not only individual feature effects, but also collective structural patterns that emerge as graph generation parameters vary. The main objectives of this work are threefold: i) to analyze how changes in graph generation parameters reshape the landscape of graph instances in feature space; ii) to investigate whether similarity in landscape (feature) space is reflected in the performance space of algorithms; and iii) to examine whether instances originating from different benchmark suites form shared or distinct landscape regions when analyzed jointly. By addressing these objectives, this study moves beyond traditional benchmarking and contributes to a more principled, structure-aware understanding of algorithm performance on graph problems, with the case-study illustrated on the notable shortest path problem, with direct relevance for algorithm analysis and selection. The data and code required to reproduce the study are available at: \url{https://zenodo.org/records/18427943}.
 
\section{Related work}
\label{related-work}
Research on graph algorithms has produced a large body of experimental studies comparing exact methods for shortest-path and related problems. Classical work focused on theoretical properties and empirical comparisons of algorithms such as Dijkstra's, Bellman–Ford, and Floyd–Warshall on synthetic graphs, often limited to small or medium instance sizes~\citep{wang2018comparison,golden1976shortest}. Subsequent research broadened these assessments to encompass larger graphs and more specialized algorithmic variants, such as bidirectional and heuristic-based methods, and was also applied to real-world road networks~\citep{gonzalez2022shortest,madkour2017survey}. This line of work has given rise to a wide range of practical applications explored in numerous studies, for example, in~\citep{alzalloum2010application,zhu2013shortest,cambazard2009shortest}.
 
Beyond direct runtime comparisons, several studies have investigated the role of structural graph characteristics in shaping algorithm behavior. Commonly analyzed properties include graph size, density, degree statistics, and edge-weight distributions. These features have been used in supervised learning settings to predict algorithm runtime or to support algorithm selection decisions~\citep{smith2012measuring}. Such approaches provide valuable insights, but typically operate at the level of individual instances.
 
Explainable AI techniques have recently been adopted to improve the interpretability of feature-based performance models. Feature attribution methods, such as SHAP, enable identification of the most influential graph characteristics for specific algorithms, offering a more nuanced understanding of algorithm performance than aggregate statistics alone~\citep{djukanovic2025learning}. However, these methods remain largely focused on feature importance, rather than relationships among instances.
 
In parallel, the evolutionary computation and optimization communities have developed instance space and landscape analysis as a framework for studying problem distributions and algorithm performance across regions of structurally similar instances. While this paradigm has proven effective for combinatorial and continuous optimization problems, its application to graph benchmarking remains limited. The present work connects these research directions by applying landscape-based analysis to graph benchmark suites, focusing on instance similarity, clustering in feature space, and the correspondence between structural landscapes and algorithm performance.
 
\section{Methodology}
\label{methodology}
 
Let $\mathcal{G}=\{g_1,g_2,\ldots,g_n\}$ denote a set of graph instances generated from one or more benchmark suites, where each instance $g_i$ is produced using a parameter configuration $\theta_j$ from a finite set of configurations $\Theta$. Each graph instance is mapped to a structural feature vector $\mathbf{x}_i \in \mathbb{R}^d$ through a feature extraction function $\phi \colon \mathcal{G}\rightarrow\mathbb{R}^d$, yielding a feature matrix $\mathbf{X}=(\mathbf{x}_1,\ldots,\mathbf{x}_n) \in \mathbb{R}^{d \times n}$. This mapping induces an instance landscape in feature space, in which structurally similar graphs are expected to be located in close proximity.
 
To identify regions of structural similarity within the instance landscape, an unsupervised clustering function $\mathcal{C}\colon \mathbb{R}^d\rightarrow\{1,\ldots,k\}$ is applied, assigning each instance, represented by a structural feature, to a cluster $c_i=\mathcal{C}(\mathbf{x}_i)$. Thus, each cluster represents a subset of graph instances sharing similar structural characteristics. The resulting partitioning of $\mathcal{G}$ defines landscape regions that are analyzed independently.
 
Each cluster is further characterized by the distribution of graph-generation parameter configurations that give rise to the instances it contains. For a cluster $c$ and the parameter configuration $\theta\in\Theta$, let $N(c,\theta)$ denote the number of instances in cluster $c$ generated using $\theta$. The cluster representation is defined as a normalized distribution over parameter configurations.
 
To assess the robustness of the identified landscape structure with respect to feature selection, alternative feature mappings are considered by applying feature filtering and selection techniques. Clustering is repeated using reduced feature sets, and the correspondence between clusters obtained from the full and reduced feature spaces is quantified using similarity measures computed on the parameter-distribution representations. High similarity indicates stable landscape regions that are invariant to the chosen feature portfolio.
 
Algorithmic performance is analyzed in a separate performance space. Let $\mathcal{A}=\{a_1,\ldots,a_m\}$ denote a set of shortest-path algorithms, and let $f_j(g_i)$ represent the runtime required by algorithm $a_j$ to solve instance $g_i$. For each cluster and algorithm, the collection of runtime values induces a performance distribution conditioned on the corresponding landscape region. To investigate whether structural similarity in the instance landscape leads to similar algorithmic behavior, non-parametric distribution statistical tests are applied to compare performance distributions across clusters and across instance subspaces defined by parameter configurations. This framework enables a principled analysis of the correspondence between structural landscapes and algorithm performance.
 
\section{Experimental design}
\label{sec:exp_design}
 
\noindent\textbf{Graph benchmark suites:} To ensure structural diversity and broad coverage of graph characteristics, we consider three benchmark suites. Each suite is defined by a distinct graph generation mechanism and parameterization, enabling controlled exploration of how structural properties evolve. \textbf{\textit{Random Graph}} is a benchmark suite consists of weighted Erdős–Rényi graphs~\citep{erdHos2013spectral}, where the edges are sampled independently with fixed probability. The primary parameters of this generator are the number of vertices and the edge density. Random graphs are widely used in algorithm analysis due to their simplicity and analytical tractability~\citep{goldberg2005computing}. From a landscape perspective, they provide a baseline setting in which structural variation is driven mainly by global parameters such as scale and density, resulting in relatively homogeneous connectivity patterns. Random graphs are generated using the Erd\H{o}s--R\'enyi model via the \texttt{NetworkX}~\citep{hagberg2020networkx} function \texttt{erdos\_renyi\_graph} with $|V|\in\{500,1000,2000,5000,10000\}$, edge probability $p\in\{0.2,0.3,0.5\}$, while edge weights are generated from $U({1, \ldots, 15})$; ten independent instances per parameter combination are constructed. \textbf{\textit{Wireless (Geometric) Graph}} is a benchmark suite based on random geometric graphs generated in a continuous metric space~\citep{kenniche2010random}. Vertices are embedded in a two-dimensional domain, and edges are added between vertices whose Euclidean distance is below a given threshold. Such graphs are commonly used to model wireless and sensor networks and introduce strong geometric and locality constraints~\citep{penrose2003random}. Compared to random graphs, wireless graphs exhibit bounded degrees, spatial clustering, and connectivity patterns driven by geometry rather than randomness. These properties make them well-suited for studying how geometric constraints influence the structure of the instance landscape. Wireless graphs are generated using the random geometric graph model via the \texttt{NetworkX} function \texttt{random\_geometric\_graph}, where vertices are placed uniformly in the unit square and edges are added between vertices within a connection radius $r\in\{0.2,0.3,0.5\}$, using the same vertex counts and repetitions; edge weights correspond to Euclidean distances. \textbf{\textit{Real-World Graph}} is a benchmark suite consists of road-network graphs derived from real geographic data. These graphs are typically sparse, near-planar, and highly structured due to physical and infrastructural constraints. Real-world graphs have been shown to induce markedly different algorithmic behavior compared to synthetic benchmarks, often favoring certain algorithmic paradigms despite similar problem sizes~\citep{goldberg2005computing,sanders2006engineering}. Their inclusion enables us to assess whether landscape structures observed in synthetic benchmarks generalize to realistic settings or remain benchmark-specific. Real-world graphs consist of road-network instances extracted using the \texttt{OSMnx}~\citep{boeing2017osmnx}, including those urban networks generated within a fixed $5\,\mathrm{km}$ radius around city centers, as well as large regional road networks--retrieving streets accessible to motor vehicles. All real-world graphs are undirected, weighted by geographic distance, and represented in \texttt{NetworkX} format. The dataset includes 25 European city graphs, with sizes ranging from 2,443 to 16,223 vertices and from 3,411 to 21,232 edges, reflecting sparse structures. In addition, 10 U.S. regional road networks\footnote{Available at~\url{http://www.diag.uniroma1.it//challenge9/download.shtml}} are involved, whose graph sizes range from 321,270 to 6,262,104 vertices and from 794,830 to 15,119,284 edges.
 
\noindent\textbf{Instance landscape features:} Each graph instance is represented by 17 structural features: number of vertices, number of edges, graph density, average vertex degree, average vertex strength, average edge weight, assortativity coefficient, minimum degree, maximum degree, standard deviation of degrees of vertices, degree of source node, the number of different nodes reachable within two hops from source node, degree of goal node, the number of different nodes that can reach goal node within two hops, minimum edge-weight, maximum edge-weight, and standard deviation of weights of edges. All features are extracted using the \texttt{NetworkX} library and computed once per instance. All features are normalized before applying clustering to ensure comparability. These attributes were selected as computational complexity of their calculation is significantly lower than the complexity of theoretically the fastest shortest-path algorithm. In other words, they are reasonably cheap to calculate.
 
\noindent\textbf{Clustering:} The \emph{k}-means algorithm is used to cluster the instances of each dataset. An appropriate number of clusters $k$ is determined using the Silhouette coefficient~\citep{shahapure2020cluster}. After clustering, the resulting clusters are analyzed through low-dimensional projections of the feature space obtained using \emph{t}-Distributed Stochastic Neighbor Embedding (t-SNE)~\citep{seo2004rank}. In addition, a coverage matrix is constructed for each benchmark suite, where each row represents instances generated using the same parameter combination, and each column corresponds to a cluster. Each cell contains the ratio of instances generated from a given parameter combination that belong to a specific cluster. By visualizing the coverage matrices, recurring structural patterns are identified, indicating that instances within the same cluster share similar structural characteristics; that is, the clusters exhibit homogeneous feature profiles.
 
\noindent\textbf{Sensitivity analysis:} In addition, to identify a subset of features that is most informative for clustering, feature selection techniques are applied to each dataset independently. Specifically, low-variance feature filtering, correlation-based feature pruning, and mutual information (MI) criteria are employed~\citep{hall1999correlation,saputra2025performance}. Using the reduced feature sets, \emph{k}-means~\citep{likas2003global} clustering and the corresponding coverage matrices are recalculated to facilitate cluster interpretation and comparison. To compare clusters obtained from the full feature set with those derived from the reduced feature sets, a cluster correspondence measure based on cosine similarity is introduced. Pairwise similarities are reported, where high cosine similarity between a pair of clusters, one obtained from the full feature set and one from the reduced feature set, indicates that they contain a highly similar subset of instances. This calculation is possible because each cluster, regardless of the feature subset used for clustering, is represented by the distribution of instances generated from the same parameter combinations within the benchmark suite.
 
\noindent\textbf{Performance analysis:}
To assess whether similar landscape regions in the feature space, as identified by the clustering results, lead to similar behavior in the performance space, non-parametric statistical tests are performed on the performance of four selected algorithms. The performance of the shortest-path algorithms is measured in terms of the runtime (wall-clock time) required to compute the (optimal) shortest path for a given graph instance. The performance analysis considers four representative shortest-path algorithms with different search and exploration characteristics. A$^*$ search~\citep{hart1968formal} is an informed search algorithm that guides exploration using a heuristic estimate of the remaining distance to the target, typically resulting in reduced search effort when the heuristic is admissible and informative. Dijkstra's algorithm is an uninformed, exhaustive method that guarantees optimal shortest paths by systematically expanding nodes in order of increasing path cost, serving as a robust baseline. Bidirectional Dijkstra accelerates the shortest-path computation by simultaneously searching forward from the source and backward from the target, often reducing the explored search space compared to the basic variant. Finally, DEQ represents a deque-based shortest-path strategy that exploits specific graph properties to improve traversal efficiency, particularly in structured or constrained graph instances. Together, these algorithms capture a spectrum of heuristic-driven and exhaustive search behaviors, enabling a comprehensive comparison of performance across different landscape regions.
 
By identifying different parameter combinations used to generate graph instances that belong to the same cluster within each benchmark dataset, the performance of each algorithm is further analyzed separately by considering the corresponding instance subspaces. Two non-parametric statistical analyses are performed. The two-sample Anderson--Darling and Kolmogorov--Smirnov tests are applied to assess whether the corresponding samples originate from the same distribution, providing a global evaluation of potential distributional shifts in the runtime performance metric across similar landscape regions. Pairwise comparisons are performed between groups within the same cluster. We are not reporting multiple pairwise comparisons, so we keep the significance level $\alpha = 0.05$.
 
\section{Results}
\label{sec:results}
 
\subsection{Impact of Graph Instance Generation Parameters on Feature-Space Landscapes}
In the first analysis, we investigate how variations in graph generation parameters affect the resulting feature-space landscape of graph instances. This analysis is conducted separately for each of the three benchmark suites. Table~\ref{tab:clustering_feature_selection} summarizes the clustering outcomes across the Random, Geometric, and Real benchmarks under different feature selection strategies.
Across all benchmarks, feature selection has a clear impact on both cluster separability and structure, as reflected by changes in the Silhouette score and the resulting number of clusters.
In particular, uncorrelated and low-variance feature subsets generally improve clustering quality compared to using all features, although the magnitude of improvement varies depending on the benchmark. We first discuss the results obtained using all features, followed by a sensitivity analysis with respect to the feature portfolio.
 
\begin{table}[t]
\centering
\caption{Clustering performance across benchmarks and feature selection strategies.}
\label{tab:clustering_feature_selection}
\small
\begin{tabular}{llccc}
\toprule
Benchmark & Feature selection & \# of features & Silhouette & \# of clusters \\
\midrule
Random & Original (All) & 17 & 0.60 & 9 \\
Random & Uncorrelated & 8 & 0.70 & 10 \\
Random & Low-variance & 13 & 0.90 & 12 \\
\midrule
Geometric & Original (All) & 17 & 0.63 & 9 \\
Geometric & Uncorrelated & 4 & 0.675 & 12 \\
Geometric & Low-variance & 14 & 0.72 & 9 \\
\midrule
Real & Original (All) & 17 & 0.30 & 5 \\
Real & Uncorrelated & 11 & 0.35 & 7 \\
Real & Low-variance & 15 & 0.42 & 5 \\
\bottomrule
\end{tabular}
\end{table}
 
\begin{figure}[t]
    \centering
    \includegraphics[width=\linewidth]{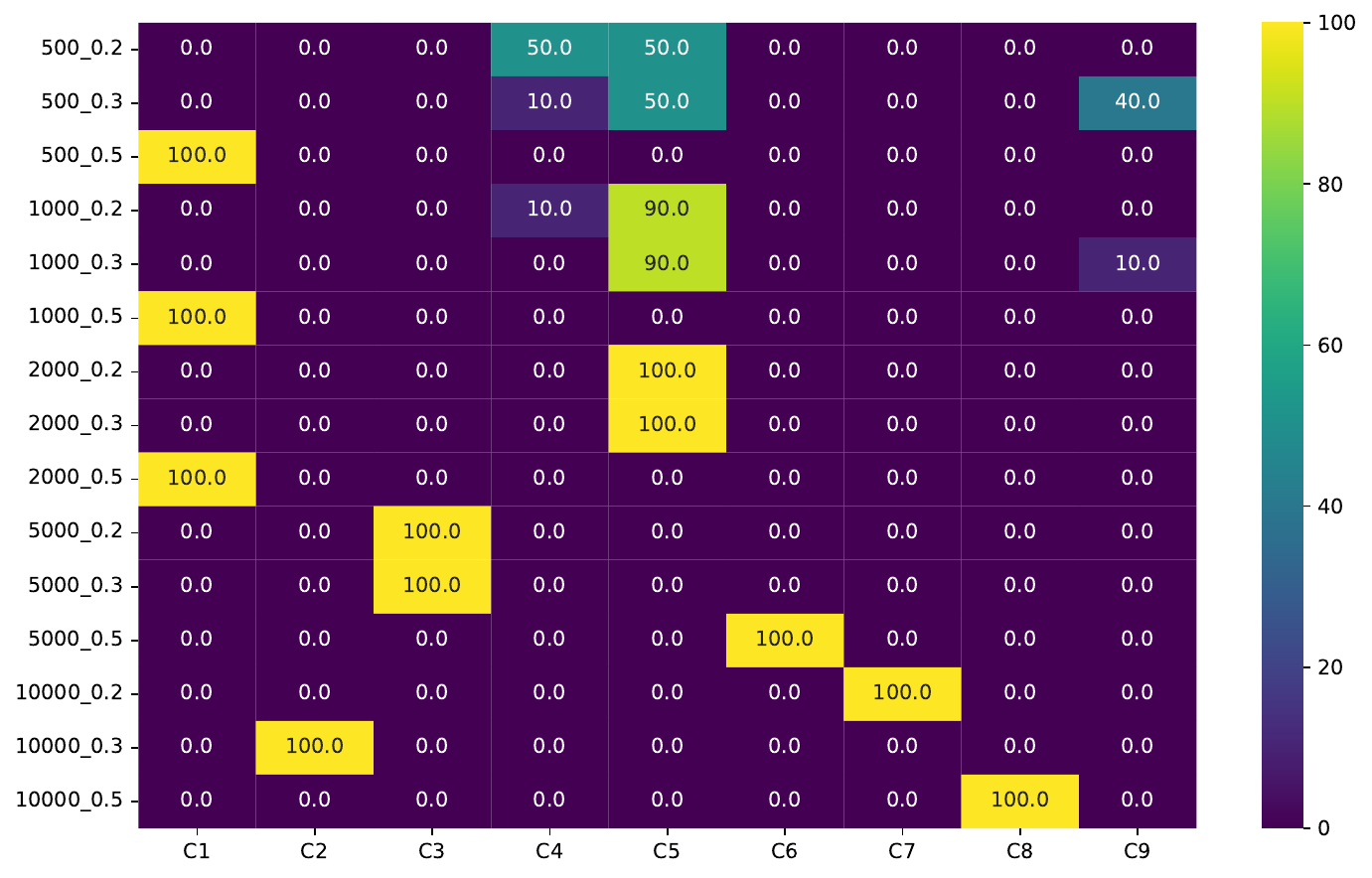}
    \caption{Coverage Matrix of Feature-Space Clusters Across Graph Generation Parameters of Random Graph Instances}
    \label{fig:Random_all_features}
\end{figure}
 
\noindent\textbf{Random graphs:} Graph instances are generated using the Erdős–Rényi model, yielding a total of 150 graphs, corresponding to distinct $(|V|, p)$ parameter combinations, with ten independent instances per combination. Each graph instance is represented using 17 features, as described in the experimental design. The instances are clustered using k-means clustering, where the number of clusters is determined based on the silhouette score, yielding an optimal value of $k=9$ with an average silhouette score of approximately 0.60. Next, we construct a coverage matrix in which rows correspond to the parameter combinations used for graph generation and columns correspond to the identified clusters. Each matrix entry represents the percentage of instances generated with the same parameter combination that are assigned to a given cluster.
 
The coverage matrix in Fig.~\ref{fig:Random_all_features} demonstrates a strong alignment between the graph generation parameters and the resulting feature-space clusters. For most $(|V|, p)$ combinations, all instances are assigned to a single cluster~[100\%], indicating that the proposed 17-dimensional feature representation consistently captures the structural properties induced by the generation process. In particular, graphs with larger numbers of nodes ($|V|\geq 2000$) or extremely dense graphs ($p=0.5$) exhibit highly stable feature profiles and form well-separated regions in feature space, as reflected by the near block-diagonal structure of the coverage matrix and an average silhouette score of approximately~[0.60].
 
Beyond parameter-wise consistency, the analysis also reveals systematic cases where different parameter combinations collapse into the same cluster, indicating feature-space equivalence across distinct generation settings. Dense graphs with $p=0.5$ and $|V|\in\{500,1000,2000\}$ are consistently grouped into a single cluster~[100\%], suggesting that a high edge probability dominates the extracted features and renders moderate variations in graph size indistinguishable after normalization. A similar trend is observed for sparser regimes with $p\in\{0.2,0.3\}$, where graphs with $|V|\in\{1000,2000\}$ are assigned to the same clusters~[90\%--100\%]. In these regimes, edge probability again acts as the primary organizing factor of the feature-space landscape, while graph size becomes discriminative only beyond certain scale thresholds. Overall, these results highlight non-uniform interactions between graph generation parameters, revealing regions of redundancy and separability that are critical for understanding benchmark diversity and coverage.
 
\begin{figure}[t]
    \centering
    \includegraphics[width=\linewidth]{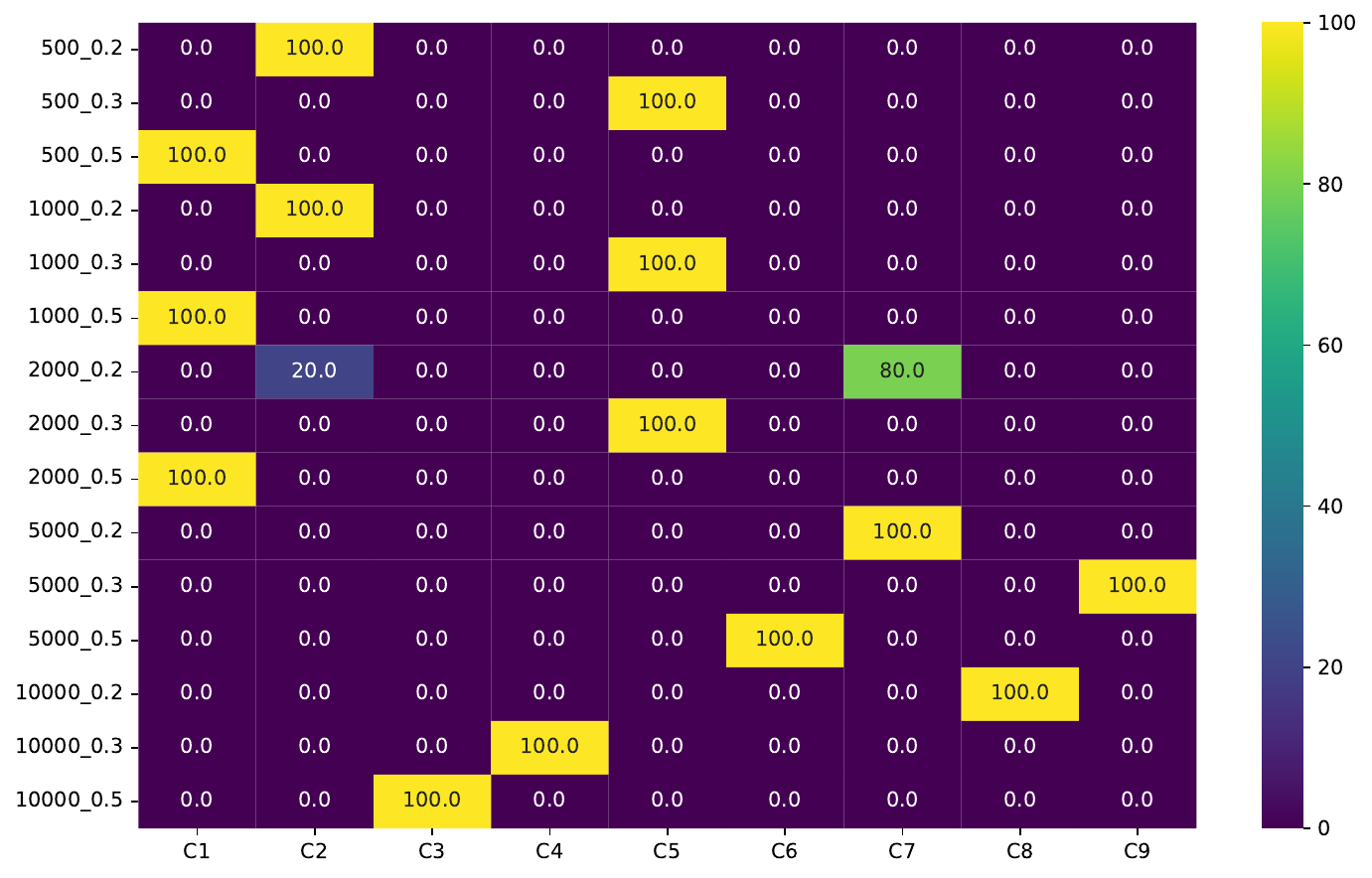}
    \caption{Coverage Matrix of Feature-Space Clusters Across Graph Generation Parameters of Geometric Graph Instances}
    \label{fig:Geometric_all_features}
\end{figure}
 
\noindent\textbf{Geometric graphs:} The coverage matrix for the geometric benchmark in Fig.~\ref{fig:Geometric_all_features} reveals a clear parameter-driven organization of the feature-space landscape, with the connection radius $r$ acting as the dominant structuring factor. In particular, dense geometric graphs with $r=0.5$ and $|V|\in\{500,1000,2000\}$ consistently collapse into the same cluster~[100\%], indicating that within this regime variations in graph size do not substantially affect the extracted feature profiles. The high connection radius induces uniformly dense local neighborhoods, leading to structurally similar graphs after feature normalization.
 
A similar size-invariant behavior is observed for sparser regimes. Graphs generated with $r=0.2$ and $|V|\in\{500,1000\}$, as well as graphs with $r=0.3$ and $|V|\in\{500,1000,2000\}$, are assigned to the same clusters~[100\%], demonstrating that spatial density dominates the feature-space representation in these ranges. Only for larger graphs or transitional connectivity settings (e.g., $|V|=2000$ with $r=0.2$) does partial cluster mixing emerge~[80\%/20\%], reflecting increased sensitivity to geometric variability at scale. Overall, these patterns confirm that for random geometric graphs, the connection radius largely determines feature-space similarity, while graph size becomes discriminative only beyond specific density and scale thresholds.
 
\noindent\textbf{Real graphs:} The experimental evaluation has been conducted on the real benchmark suite comprising 35 real-world road network instances, which exhibit substantial heterogeneity in problem scale. To systematically analyze the instance characteristics, we have partitioned the dataset based on quartile intervals of the number of nodes. The instances range from 2,443 to 6,262,104 nodes, with quartile boundaries at 5,560 (Q1), 7,735 (Q2), and 292,808 (Q3) nodes. The first quartile (Q1: 2,443–5,560 nodes) contains 9 instances representing small-to-medium European urban networks, including Kyiv, Helsinki, Minsk, Ljubljana, Bratislava, Bern, Moscow, Oslo, and Warsaw. The second quartile (Q2: 5,560–7,735 nodes) comprises 9 medium-sized European city networks: Berlin, Belgrade, Zagreb, Budapest, Stockholm, Sofia, Lisbon, Copenhagen, and Vienna. The third quartile (Q3: 7,735–292,808 nodes) includes 8 instances spanning larger European capitals and one USA regional network: Dublin, Brussels, Bucharest, Paris, Rome, Madrid, London, and USA-NY. The fourth quartile (Q4: 292,808–6,262,104 nodes) consists of 9 large-scale US road networks from the 9th DIMACS Implementation Challenge: BAY, CAL, COL, E, FLA, LKS, NE, NW, and W. This stratification reveals a clear demarcation between European urban networks, which predominantly occupy the lower three quartiles, and the substantially larger US regional road networks concentrated in the upper quartile, thereby enabling comprehensive evaluation of algorithmic scalability across four orders of magnitude in problem size.
 
\begin{figure}[t]
    \centering
    \includegraphics[width=\linewidth]{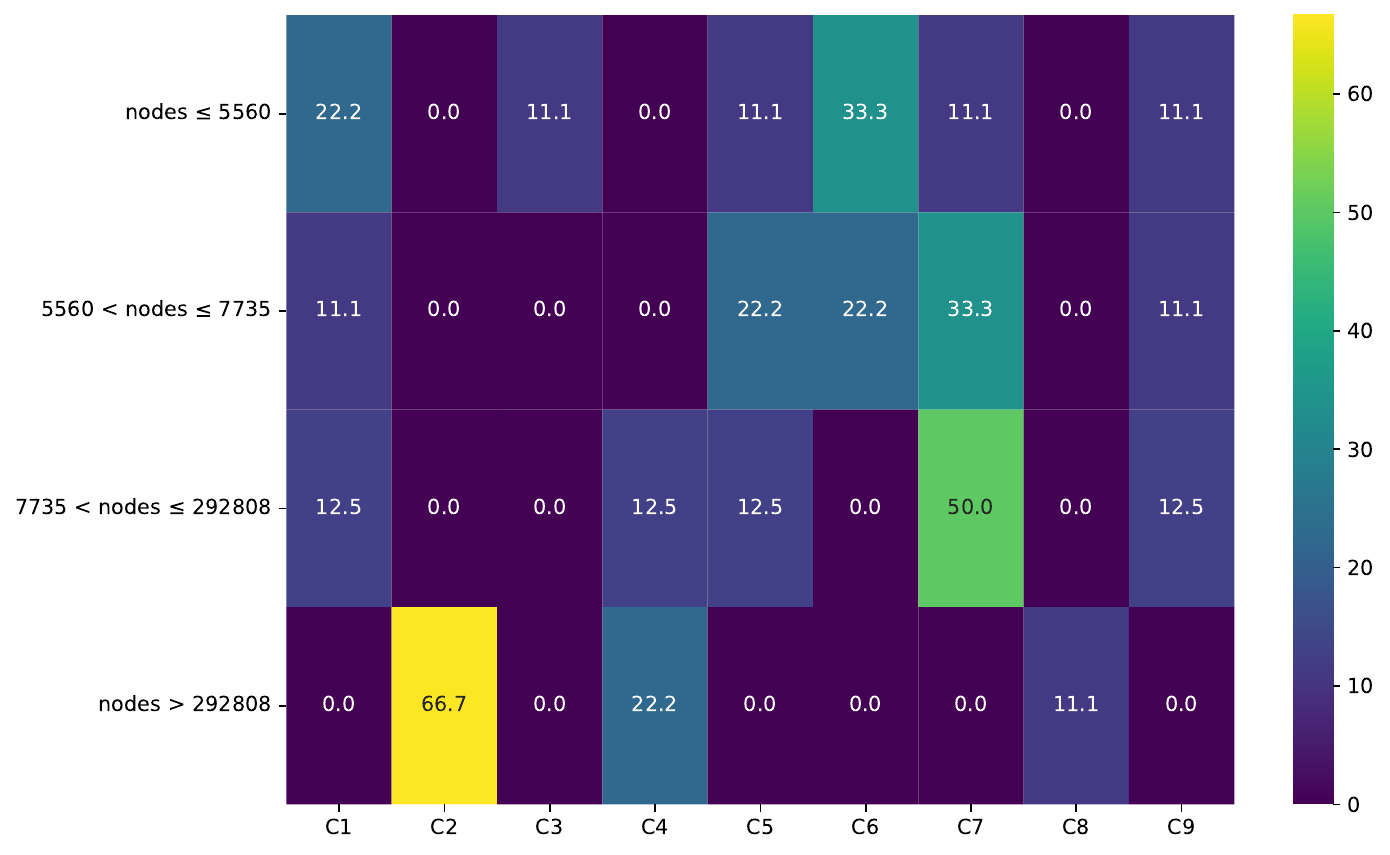}
    \caption{Coverage Matrix of Feature-Space Clusters Across Graph Generation Parameters of Real Graph Instances.}
    \label{fig:Real_all_features}
\end{figure}
 
To investigate the relationship between instance scale and structural characteristics, we analyzed the distribution of instances across the nine identified clusters (C1–C9) stratified by node-count quartiles. Fig.~\ref{fig:Real_all_features} presents a heatmap illustrating the percentage of instances from each quartile interval assigned to each cluster. The results reveal distinct clustering patterns that correlate with problem size. Small-scale instances (Q1: $|V| \leq$ 5,560) exhibit a dispersed distribution, with the highest concentration in cluster C6 (33.3\%), followed by C1 (22.2\%). Medium-sized instances (Q2: 5,560 $< |V| \leq$ 7,735) are predominantly assigned to clusters C7 (33.3\%), C5 (22.2\%), and C6 (22.2\%), indicating structural similarities among European urban networks of comparable size. The third quartile (Q3: 7,735 $< |V| \leq$ 292,808) shows a strong affinity toward cluster C7, which captures 50.0\% of instances in this range. Most notably, large-scale instances (Q4: $|V|>$ 292,808) demonstrate a markedly different clustering behavior, with 66.7\% concentrated in cluster C2 and 22.2\% in C4, while exhibiting zero membership in clusters C1, C3, C5, C6, C7, and C9. This pronounced separation suggests that US road networks possess fundamentally different properties compared to European urban networks, which the clustering algorithm successfully captures. The absence of overlap between Q4 instances and the clusters dominated by smaller networks (C5, C6, C7) confirms that problem scale is strongly associated with distinct graph characteristics beyond mere size differences.
 
\noindent\textbf{Feature-Portfolio Sensitivity of Clustering Results:} To assess sensitivity to the feature portfolio, we compare clustering results from the full feature set with those from feature-selected subsets for each benchmark. Clusters are contrasted using coverage matrices that capture the distribution of instances per parameter combination. This allows us to identify clustering structures that remain stable under feature reduction and those that change when redundant or less informative features are removed.
 
\begin{figure}[t]
    \centering
    \includegraphics[width=\linewidth]{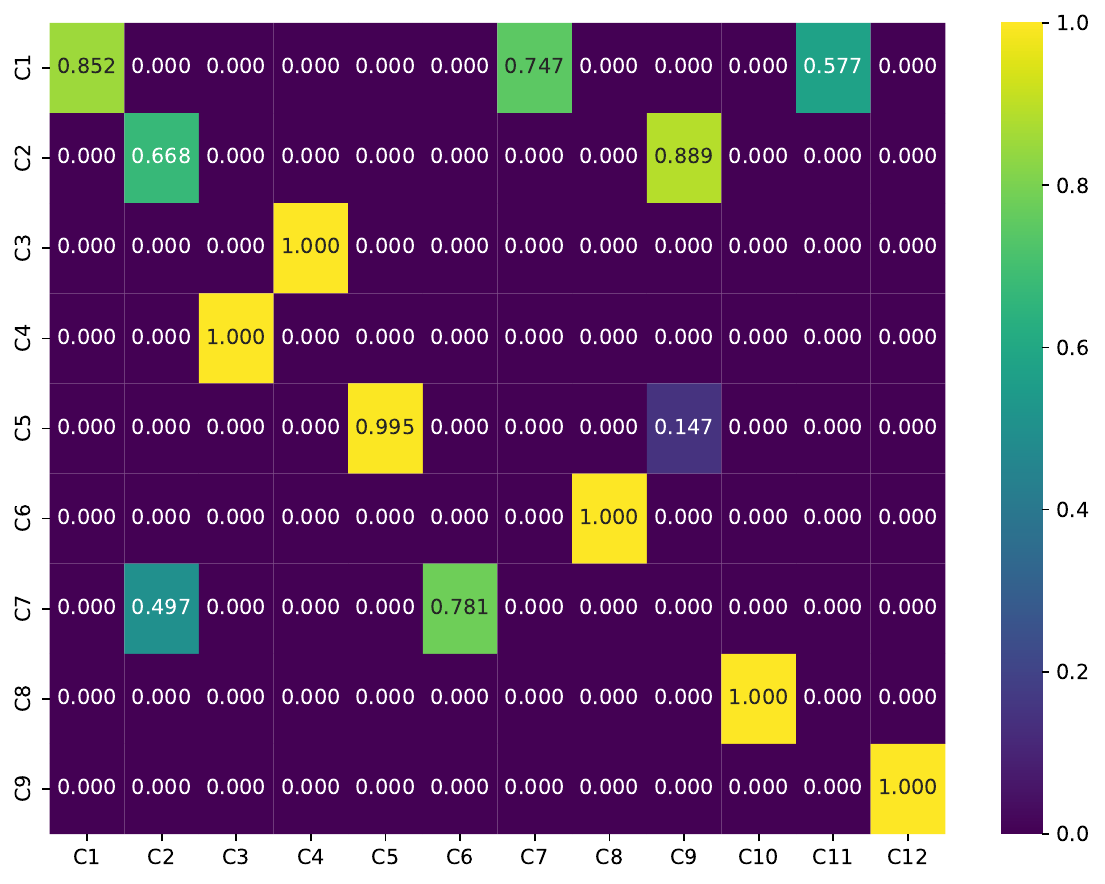}
    \caption{Cluster similarity matrix between the original feature-based clustering (k = 9) and the clustering obtained using the uncorrelated feature subset (k = 12) for the geometric benchmark.}
    \label{fig:geometric_all_uncorrelated}
\end{figure}
 
\noindent\textbf{Geometric benchmark:} Within the geometric benchmark, across both feature-selection strategies (uncorrelated vs. all features), the heatmaps show largely stable clustering structures, with most clusters preserving strong one-to-one correspondences and indicating that parameter-driven patterns are robust to feature reduction (see Fig.~\ref{fig:geometric_all_uncorrelated}, low-variance vs. all features heatmap is in our repository). While correlation-based feature removal introduces slightly more off-diagonal overlap than low-variance filtering, these deviations are localized and do not lead to substantial changes in the overall clustering outcome.
 
\noindent\textbf{Random benchmark:} For the random benchmark, the coverage matrices reveal noticeably lower clustering stability under feature reduction compared to the geometric benchmark, with several clusters fragmenting or redistributing their mass across multiple counterparts; however, they still provide stable results in general. Both low-variance and uncorrelated feature selection introduce substantial off-diagonal structure, indicating that clustering outcomes are more sensitive to the chosen feature portfolio and less dominated by parameter-driven patterns (heatmaps available in our repository). \textbf{Real benchmark:} Here, no significant differences are observed in the distribution of instances across clusters when different feature portfolios are used, which is also reflected in the silhouette scores and the number of clusters reported in Table~\ref{tab:clustering_feature_selection}.
 
\subsection{Do Similar Feature-Space Landscapes Imply Similar Algorithm Performance?}
 
To examine whether structural similarity in the instance landscape, induced by different generator parameter combinations, corresponds to similar algorithmic behavior, we apply non-parametric tests to compare algorithm performance across parameter-defined instance subspaces within the same cluster. We focus on cases where different parameter combinations lead all generated instances to the same cluster (100\% coverage), ensuring a consistent landscape definition. In these cases, performance comparisons are based on ten instances, whereas other configurations yield fewer instances and would violate the assumptions of some statistical tests. For the random benchmark, such cases include the parameter sets $(500, 0.5)$, $(1000, 0.5)$, and $(2000, 0.5)$; $(2000, 0.2)$ and $(2000, 0.3)$; and $(5000, 0.2)$ and $(5000, 0.3)$. For the four algorithms in our portfolio, we compare performance across all selected instance sets.
 
\begin{figure}[t]
    \centering
    \includegraphics[width=\linewidth]{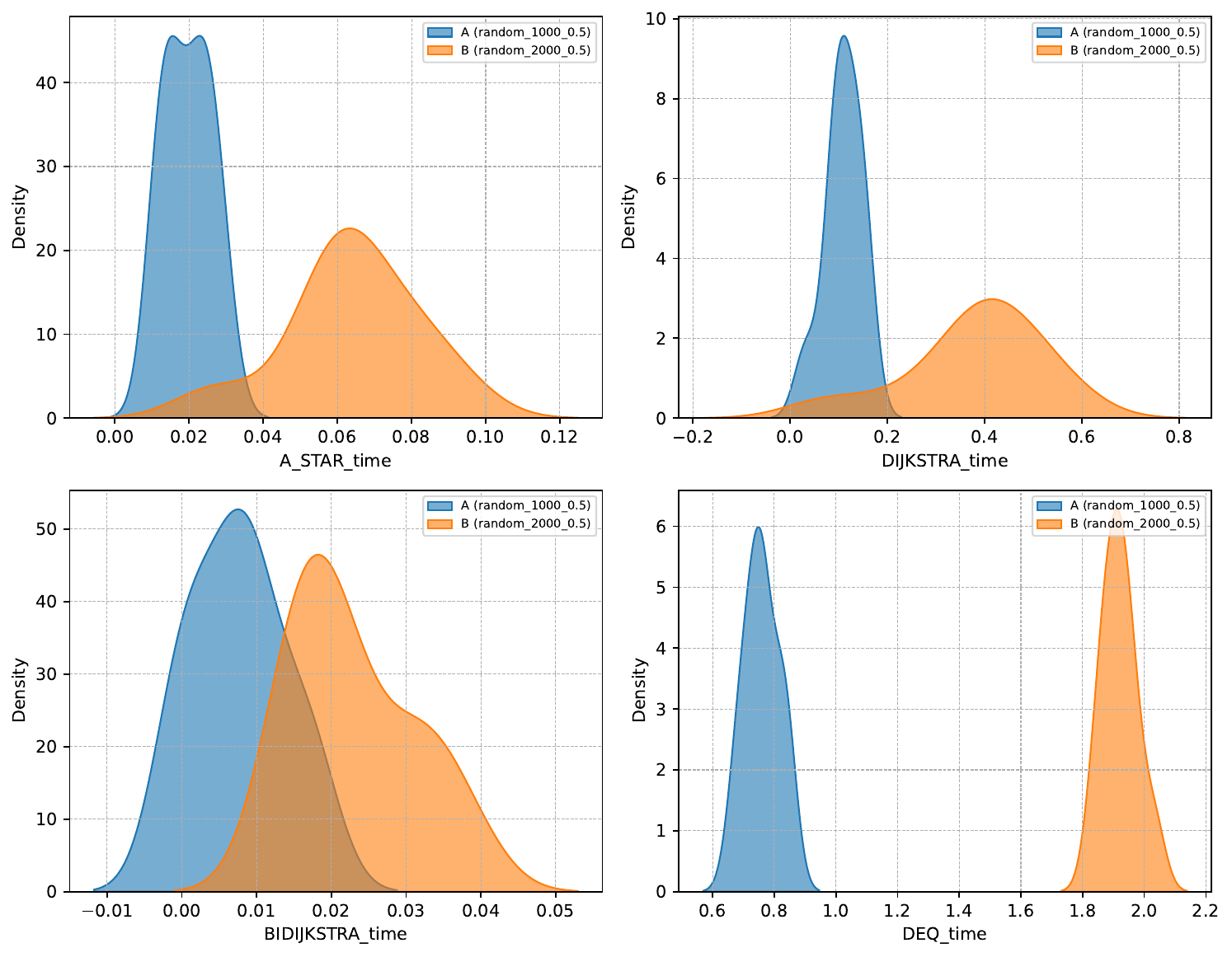}
    \caption{Kernel Density Estimates of Algorithm Performance for $(|V|, p) \in \{(1000, 0.5), (2000, 0.5)\}$ within the Random Graphs.}
    \label{fig:kde_performance}
\end{figure}

Table~\ref{tab:ad_combined} reports the Anderson--Darling (AD) test results comparing runtime distributions across blocks of random graph instances for all evaluated algorithms. The Kolmogorov--Smirnov (KS) test was additionally applied to all comparisons and is provided in the accompanying repository; both tests yield fully consistent statistical decisions, confirming the robustness of the observed outcomes~\citep{eftimov2018impact}. For A$^{*}$, no statistically significant difference is observed only for the smallest density variation ($n$ = 2000; $p$ = 0.2\ vs\ $p$ = 0.3), whereas all remaining comparisons indicate significant distributional shifts. Bidirectional Dijkstra exhibits partial robustness, with non-significant or borderline p-values for small-scale changes ($n$ = 2000; $p$ = 0.2\ vs\ $p$ = 0.3 and $n$ = 500\ vs\ $n$ = 1000; $p$ = 0.5), while larger graph variations lead to statistically significant differences. For Dijkstra, the test indicates no statistically significant difference only for the comparison ($n$ = 2000; $p$ = 0.2\ vs\ $p$ = 0.3), where the p-value exceeds 0.05. In contrast, all remaining comparisons yield p-values below the significance threshold, indicating statistically significant distributional difference. In contrast, DEQ consistently exhibits statistically significant differences across all graph families. In addition, Fig.~\ref{fig:kde_performance} illustrates the kernel density estimates of algorithm performance over these instances generated with $(1000, 0.5)$, and $(2000, 0.5)$. This result supports the statistical tests, indicating that although landscape features group these instances into the same cluster, this structural similarity is not reflected in the performance behavior of the four algorithms. Overall, the AD results, confirmed also by the KS tests, reveal algorithm-dependent sensitivity patterns, with limited robustness under mild graph variations.
 
\begin{table}[t]
\centering
\caption{Two-sample Anderson--Darling (AD) test p-values for runtime distribution comparisons across blocks of random and geometric graph instances.}
\label{tab:ad_combined}
\small
\begin{tabular}{llcccc}
\toprule
\textbf{Graph type} & \textbf{Instance blocks} & \textbf{A$^{*}$} & \textbf{DIJKSTRA} & \textbf{BIDIJKSTRA} & \textbf{DEQ} \\
\midrule
\multirow{5}{*}{Random}
& $n$ = 2000; $p$ = 0.2 vs $p$ = 0.3      & 0.250 & 0.250 & 0.067 & 0.001 \\
& $n$ = 5000; $p$ = 0.2 vs $p$ = 0.3      & 0.001 & 0.001 & 0.001 & 0.001 \\
& $n$ = 500 vs $n$ = 1000; $p$ = 0.5      & 0.004 & 0.001 & 0.076 & 0.001 \\
& $n$ = 500 vs $n$ = 2000; $p$ = 0.5      & 0.001 & 0.001 & 0.001 & 0.001 \\
& $n$ = 1000 vs $n$ = 2000; $p$ = 0.5     & 0.001 & 0.001 & 0.001 & 0.001 \\
\midrule
\multirow{7}{*}{Geometric}
& $n$ = 500 vs $n$ = 1000; $r$ = 0.2      & 0.078 & 0.001 & 0.026 & 0.001 \\
& $n$ = 500 vs $n$ = 1000; $r$ = 0.3      & 0.250 & 0.018 & 0.018 & 0.001 \\
& $n$ = 500 vs $n$ = 2000; $r$ = 0.3      & 0.196 & 0.001 & 0.001 & 0.001 \\
& $n$ = 1000 vs $n$ = 2000; $r$ = 0.3     & 0.025 & 0.001 & 0.001 & 0.001 \\
& $n$ = 500 vs $n$ = 1000; $r$ = 0.5      & 0.250 & 0.001 & 0.006 & 0.001 \\
& $n$ = 500 vs $n$ = 2000; $r$ = 0.5      & 0.009 & 0.001 & 0.001 & 0.001 \\
& $n$ = 1000 vs $n$ = 2000; $r$ = 0.5     & 0.018 & 0.001 & 0.001 & 0.001 \\
\bottomrule
\end{tabular}
\end{table}
 
For the geometric benchmark, the performance analysis is conducted on selected instance sets corresponding to parameter combinations that fully collapse into the same cluster, namely $(500, 0.5)$, $(1000, 0.5)$, and $(2000, 0.5)$; $(500, 0.2)$ and $(1000, 0.2)$; and $(500, 0.3)$, $(1000, 0.3)$, and $(2000, 0.3)$. Table~\ref{tab:ad_combined} reports the Anderson--Darling (AD) test results for runtime distribution comparisons across geometric graph families. The Kolmogorov--Smirnov (KS) test was additionally applied to all comparisons and is provided in the accompanying repository; both tests yield fully consistent statistical decisions. For A$^{*}$, no statistically significant differences are observed for most density and size variations, including $n$ = 2000; $p$ = 0.2\ vs\ $p$ = 0.3; $n$ = 500\ vs\ $n$ = 1000; $r$ = 0.3; $n$ = 500\ vs\ $n$ = 2000; $r$ = 0.3; and $n$ = 500\ vs\ $n$ = 1000; $r$ = 0.5, with statistically significant differences emerging only for larger size changes such as $n$ = 1000\ vs\ $n$ = 2000; $r$ = 0.3; $n$ = 500\ vs\ $n$ = 2000; $r$ = 0.5; and $n$ = 1000\ vs\ $n$ = 2000; $r$ = 0.5. In contrast, Bidirectional Dijkstra, Dijkstra and DEQ yield p-values below 0.05 for all evaluated geometric graph comparisons, indicating systematic sensitivity of their runtime distributions to changes in graph structure.
 
This analysis is not performed for the real benchmark because splitting by node quantiles yields too few instances per cluster for reliable statistical testing (often only two or three). As a result, distributional tests such as Anderson–Darling or Kolmogorov–Smirnov are not meaningful in this setting.
 
Overall, the results achieved on the random and geometric benchmark confirm the statistical tests and show that feature-space similarity does not imply similar algorithmic behavior. While clustering captures structural similarities, performance remains sensitive to scale effects not fully represented by the chosen features, revealing a gap between landscape similarity and performance equivalence.
 
\subsection{Cross-Benchmark Analysis of Feature-Space Landscape Regions}
 
\begin{figure}[t]
    \centering
    \includegraphics[width=\linewidth]{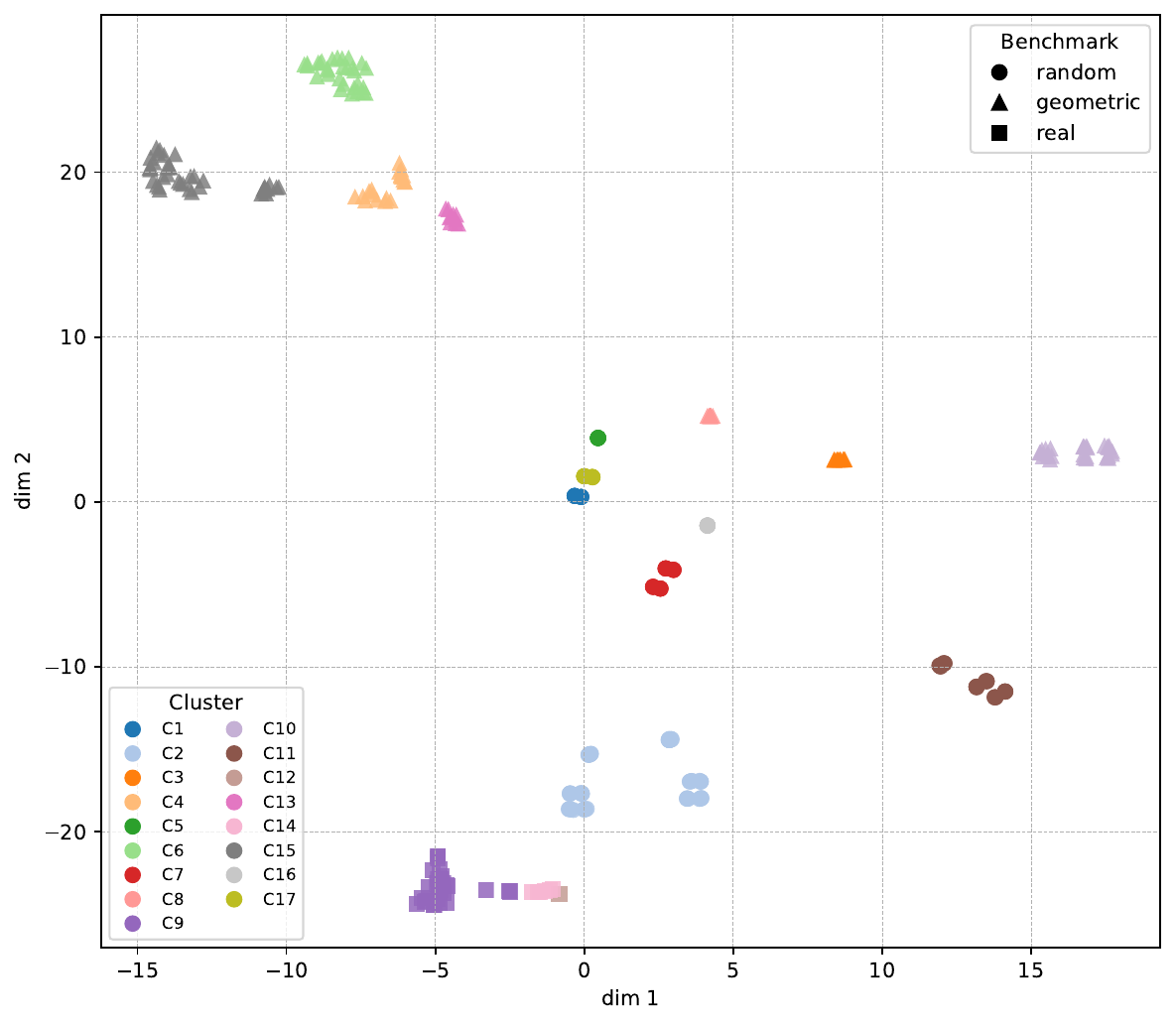}
    \caption{Two-dimensional t-SNE embedding of the merged benchmarks. Points are colored according to cluster labels and different marker shapes indicate instances origin (random, geometric, and real).}
    \label{fig:All_instances_tSNE}
\end{figure}

Fig.~\ref{fig:All_instances_tSNE} presents the t-SNE visualization of cluster composition when all benchmark suites are merged. The result shows that instances from random, geometric, and real benchmarks form well-separated groups in the low-dimensional embedding. Clusters appear compact and largely homogeneous with respect to their benchmark origin, while spatial separation between groups is pronounced. Only minimal proximity is observed at the boundaries of some clusters, but there is no substantial overlap between the benchmark families. This indicates that, when analyzed jointly, instances originating from different benchmark suites occupy distinct regions of the landscape feature space, rather than collapsing into shared regions.
 
\section{Discussion}
\label{sec:discussion}
We adopt an instance-landscape view of graph benchmarking by clustering graphs in a structural feature space and analyzing how generator parameters and benchmark origin shape the resulting regions. While parameter settings form stable clusters and geometric graphs are more robust to feature reduction than random graphs, shortest-path runtime distributions often differ significantly even within the same feature-space region (especially for Dijkstra, Bidirectional Dijkstra, and DEQ), with A$^{*}$ showing greater but not complete stability, and mixed suites occupying largely distinct regions. Key limitations are that (i) the landscape features are intentionally cheap and may not capture algorithm-relevant scale effects or fine-grained topology (e.g., planarity constraints, separator structure, hierarchy), (ii) k-means imposes spherical cluster bias and requires selecting $k$, (iii) the statistical tests rely on sufficient within-region sample size, which prevents analogous distributional testing on the real benchmark under the node-quantile split due to very small per-cluster counts, and (iv) results are based on the selected algorithm portfolio and runtime measurement setup, so conclusions should be interpreted as structural–performance associations rather than causal claims.
 
\section{Conclusion}
\label{sec:conclusion}
This work shows that graph benchmark suites induce structured, region-like instance landscapes in a low-cost structural feature space, and that these regions can be summarized and compared using coverage matrices and cluster correspondence measures. Across synthetic benchmarks, generator parameters (density/radius and scale) largely determine landscape regions, and mixed-suite analysis indicates that random, geometric, and road-network graphs occupy mostly disjoint regions, highlighting strong benchmark-specific structure. Crucially, we find that structural similarity at the landscape level does not guarantee performance equivalence: runtime distributions can shift significantly within the same landscape region, particularly for exhaustive search variants, revealing that performance remains sensitive to effects not fully represented by the chosen features. Future work will expand the feature space with topology- and road-network–specific descriptors (e.g., planarity and separator proxies, hierarchy indicators), evaluate alternative clustering models (e.g., GMM/HDBSCAN) and stability criteria, and scale the real-benchmark analysis by increasing the number of instances per regime (e.g., controlled subsampling or additional regions) to enable distributional testing. Finally, we will leverage the discovered landscape regions for region-aware algorithm selection and for designing more representative graph benchmark sets with explicit coverage guarantees.
 
\section*{Acknowledgements}
The authors acknowledge the support of the Horizon Europe EU research and innovation framework ERA Chair AutoLearn-SI (101187010), as well as the Slovenian Research Agency through program grants No. P2-0098 and No. P2-0103, project grant No. J2-70078 and No. GC-0001. This work is co-funded by the European Union's Horizon Europe research and innovation program under the Marie Sklodowska-Curie COFUND Postdoctoral Programme (grant agreement No.~101081355 -- SMASH), and by the Republic of Slovenia and the European Union through the European Regional Development Fund.

\end{document}